\documentclass[aps,prb,superscriptaddress,twocolumn,floatfix,notitlepage,a4paper]{revtex4-1}

\usepackage{graphicx}
\usepackage{amsmath,amssymb,amsfonts}
\usepackage{bm}
\usepackage{color}
\usepackage[percent]{overpic}
\usepackage{paralist}
\usepackage[breaklinks=true,colorlinks,citecolor=blue,linkcolor=blue,urlcolor=blue]{hyperref}

\begin{document}

\title{Plasmons in realistic graphene/hexagonal boron nitride moir\'e patterns}

\date{\today}

\author{Andrea Tomadin}
\email{andrea.tomadin@mac.com}
\affiliation{Istituto Italiano di Tecnologia, Graphene Labs, Via Morego 30, 16163 Genova, Italy}

\author{Marco Polini}
\affiliation{Istituto Italiano di Tecnologia, Graphene Labs, Via Morego 30, 16163 Genova, Italy}

\author{Jeil Jung}
\email{jeiljung@uos.ac.kr}
\affiliation{Department of Physics, University of Seoul, Seoul 02504, Korea}

\begin{abstract}
Van der Waals heterostructures employing graphene and hexagonal boron nitride (hBN) crystals have emerged as a promising platform for plasmonics thanks to the tunability of their collective modes with carrier density and record values for plasmonics figures of merit.
In this Article we investigate theoretically the role of moir\'e-pattern superlattices in nearly aligned graphene on hBN by using continuum-model Hamiltonians derived from {\it ab initio} calculations. We calculate the system's energy loss function for a variety of chemical potential values that are accessible in gated devices.
Our calculations reveal that the electron-hole asymmetry of the moir\'e bands leads to a remarkable asymmetry of the plasmon dispersion between positive and negative chemical potentials, showcasing the intricate band structure and rich absorption spectrum across the secondary Dirac point gap for the hole bands.
\end{abstract}

\maketitle

\section{Introduction}

Research on graphene has been actively pursued during the past decade after seminal experiments achieving its isolation via micromechanical cleavage and subsequent transport measurements.~\cite{novoselov2004electric,novoselov2005two,geim2007rise,neto2009electronic, sarma2011electronic}
Graphene shows semimetallic behavior with linearly dispersing massless Dirac-fermion bands near charge neutrality while hexagonal boron nitride (hBN) is a wide band-gap insulator.
Devices based on graphene on hBN substrates~\cite{dean2010boron,geim2013van,xue2011scanning} have shown a dramatic enhancement of electronic quality compared to ${\rm Si}{\rm O}_2$ substrates, due to the atomically smooth surface structure of hBN and the relatively smaller number of charge trap centers due to dangling bonds.
This has allowed the observation of fine Coulomb-interaction-driven phenomena such as fractional quantum Hall states,~\cite{du2009fractional,bolotin2009observation} enhancement of the Fermi velocity,~\cite{elias2011dirac} and strong Coulomb drag,~\cite{gorbachev2012strong} to mention a few examples.

The use of hBN substrates allows to engineer the electronic structure of graphene.
In the limit where the layers are nearly aligned, graphene on hBN (G/hBN) displays moir\'e superlattice patterns with periodicity as large as $ \lambda \sim 14~{\rm nm}$, which allows to access moir\'e-pattern-induced electronic structure features at carrier densities accessible with conventional electrostatic gating.
The presence of a long periodic superlattice structure together with the high quality of the devices has allowed the observation of Hofstadter butterfly physics in the presence of magnetic fields routinely accessible in the laboratory.~\cite{dean2013hofstadter,ponomarenko2013cloning}
The moir\'e periodicity $\lambda$ depends on the relative difference $\epsilon = ( a_{\rm G} - a_{\rm BN} )/ a_{\rm BN}$ between graphene's and hBN's lattice constants, $a_{\rm G}$ and $a_{\rm BN}$, respectively, and on the relative twist angle $\theta$, through $\lambda \simeq a_{\rm G} / ( \theta^2 + \epsilon^2 )^{1/2}$.
Because of this dependence, the layer orientation is an additional knob for modifying the electronic structure.
The band gaps observed at the primary~\cite{amet2013insulating,hunt2013massive, florida_2014} and secondary Dirac points~\cite{wang2016direct} reflect the effect of moir\'e strains inducing an average mass term,~\cite{woods2014commensurate,jung2015origin,sanjose2014spontaneous} while realistic models for the moir\'e pattern potentials can capture the secondary Dirac point gaps on the hole side.~\cite{dasilva2015transport, jung_prb_2017}
Fairly accurate moir\'e patterns can be modeled already in the first harmonic approximation for slowly varying potentials in the basis of the identity and Pauli matrices, whose details determine the character of the moir\'e superlattice bands such as the electron-hole asymmetry~\cite{dasilva2015transport} and the presence of secondary gaps.~\cite{jung_prb_2017}

A detailed study of the collective behavior induced by long-range Coulomb interactions in G/hBN systems is of considerable value for advancing our understanding of 2D-material-based plasmonics.
Indeed, it has been recognized early on that graphene and other two-dimensional materials exhibit very interesting optoelectronic properties.~\cite{bonaccorso_naturephoton_2010}
In particular, G/hBN systems have been identified as a promising platform for plasmonic applications,~\cite{grigorenko_naturephoton_2012} featuring e.g.~electrical tunability of the plasmon dispersion, high compression of  electromagnetic radiation, and facile coupling to emitters adjacent to the graphene sheet.
Most importantly, it has been shown~\cite{woessner_naturemater_2015} that graphene encapsulated between hBN crystals supports plasmon propagation with room-temperature lifetime $\tau_{\rm p}$ exceeding $500~{\rm fs}$, which represents a five-fold enhancement compared to that achieved in the case of ${\rm Si}{\rm O}_2$~\cite{fei_nature_2012} or ${\rm SiC}$~\cite{,chen_nature_2012} substrates. The propagation of plasmons in graphene/hBN systems has also been used to reconstruct the local conductivity,~\cite{ni_naturemater_2015} and thus verify the modification of the electronic structure due to the moir\'e pattern.
More recently, the plasmonic properties of encapsulated graphene have been explored at liquid-nitrogen temperatures, where plasmon lifetimes on the order of $1.600~{\rm fs}$ have been measured.~\cite{ni_nature_2018}

These breakthroughs have fostered a substantial research activity into the optoelectronic properties of the large family of two-dimensional materials, which includes semimetals, semiconductors, and insulators.~\cite{basov_science_2016, low_naturemater_2017}
These materials feature several light-matter hybrid modes, generally referred to as ``polaritons,'' which are supported by the electric polarization of free carriers, excitonic states, or lattice ions.

Previous studies of the electron energy loss in G/hBN moir\'e patterns,~\cite{tomadin_prb_2014} which neglected gauge fields, demonstrated that the dispersion of plasmonic excitations in graphene sensitively depends on the Hamiltonian details. In this Article we take a step further by exploring plasmons in G/hBN in a wide chemical potential range accessible in experiments by using a {\it realistic} electronic structure model for G/hBN moir\'e patterns.~\cite{jung2014ab,jung2015origin,jung_prb_2017}
The manuscript is structured as follows.
In Sec.~\ref{sec:model} we present the continuum-model effective Hamiltonian for G/hBN, summarizing the framework to calculate a set of realistic parameters for the moir\'e potential, and we detail the expression for the dielectric and loss functions within the random phase approximation (RPA).~\cite{Giuliani_and_Vignale}
The results of our numerical calculations are presented and discussed in Sec.~\ref{sec:results}.
Finally, in Sec.~\ref{sec:summary} we summarize our work and draw our main conclusions.

\section{The model}
\label{sec:model}

\subsection{Effective Hamiltonian}

The effect of interlayer coupling on graphene's band structure can be modelled through the following Hamiltonian, where we use the notation of Ref.~\onlinecite{tomadin_prb_2014}, including scalar, mass, and gauge potentials:
\begin{eqnarray}\label{eq:hamiltonian}
\hat{\cal H} & = & v_{\rm F} \tau_{0} {\bm \sigma} \cdot \hat{\bm p} + \tau_{0} \sigma_{0} V({\bm r}) + \tau_{3} \sigma_{3} \lbrack \Delta_{0} + \Delta({\bm r}) \rbrack \nonumber \\
& & + \tau_{3} {\bm \sigma} \cdot {\bm A}({\bm r})~.
\end{eqnarray}
The position vector ${\bm r}$ lies in the two-dimensional (2D) plane where the graphene sheet lies.
The Pauli matrices $\sigma_{i}$ operate on the pseudospin space spanned by the sublattice sites $A$, $B$; $\tau_{i}$ acts on the space of graphene's principal valleys $K$, $K'$; and ${\bm \sigma} = (\sigma_{1}, \sigma_{2})$ is a vector of Pauli matrices.
We parametrize the spatial distribution of the moir\'e pattern as follows~\cite{jung_prb_2017,footnote_phi}
\begin{eqnarray}\label{H0}
V({\bm r}) & = & 2 C_{0} \Re e [e^{i \phi_{0}} \hspace{-1pt} f({\bm r})]~,  \\
\label{Hz}
\Delta({\bm r}) & = & 2 C_{z} \Re e [e^{i \phi_{z}} \hspace{-1pt} f({\bm r})]~,  \\
{\bm A}({\bm r}) & = & 2 C_{AB} \chi \hat{z} \times \vec{\nabla}\,\Re e[e^{i \phi_{xy}} f({\bm r})]~.
\label{Hxy}
\end{eqnarray}
Here, $C_{\mu}$ and $\phi_{\mu}$ are numerical constants that we discuss below.
The coefficient $\chi$ depends on the twist angle $\theta$ and is given by
\begin{equation}
\chi = \frac{1+ \epsilon - \cos{(\theta)}}{\sqrt{(1+\epsilon)^{2} + 2 (1 + \epsilon) \cos{(\theta)} + 1}}~.
\end{equation}
Finally, the complex function $f({\bm r})$ is given by
\begin{equation}
f({\bm r}) = \sum_{m = 1}^{6} e^{i {\bm G}_{m} \cdot {\bm r}} \frac{1 + (-1)^{m}}{2}~,
\end{equation}
where ${\bm G}_{m}$ are the six reciprocal lattice vectors of the moir\'e pattern which are closer to the origin.
The Hamiltonian~(\ref{eq:hamiltonian}) can be rewritten in the more explicit form
\begin{equation}
\label{moireeq}
\hat{\cal H} = v_{\rm F} \tau_{0} {\bm \sigma} \cdot \hat{\bm p} + \tau_{3} \sigma_{3} \Delta_{0} + \sum_{m = 1}^{6} e^{i {\bm G}_{m} \cdot {\bm r}} W_{m}~,
\end{equation}
where
\begin{widetext}
\begin{equation}
\label{wexpression}
W_{m} = \left(
\begin{array}{cc}
W_{m,1}  & 0 \\
0  &  W_{m,-1}
\end{array}
\right), \quad
W_{m,\nu} =
\left(
\begin{array}{cc}
C_0 e^{i (-1)^m \phi_0} + \nu C_z e^{i (-1)^m \phi_z}    &   \nu C_{AB} \chi e^{i [ - \varphi_m + (-1)^{m+1} \phi_{xy} ]}  \\
- \nu C_{AB} \chi e^{i [ \varphi_m + (-1)^{m+1}  \phi_{xy} ]} &  C_0 e^{i (-1)^m \phi_0} - \nu C_z e^{i (-1)^m \phi_z}   \\
\end{array}
\right)~.
\end{equation}
\end{widetext}
Here, $\varphi_{m}$ is the polar angle of the wave vector ${\bm G}_{m}$ and the order of the basis vectors in the $4$-dimensional sublattice/valley space is $\{|AK\rangle$, $|BK\rangle$, $|BK'\rangle$, $-|AK'\rangle\}$.

The Hamiltonian for G/hBN that we use {in this work is based on {\it ab initio} calculations for the interlayer coupling}, which capture effects beyond the commonly assumed two-center approximation.~\cite{jung2014ab}
In Eq.~(\ref{H0}), the magnitude and phase pairs $C_{\mu}$, $\phi_{\mu}$ are moir\'e pattern parameters that capture the effective interlayer coupling in the first harmonic approximation.~\cite{jung2014ab,jung_prb_2017}
The parameters $C_{\mu}$, $\phi_{\mu}$ were obtained by first calculating the distant hopping terms from carbon to the substrate atoms, for all possible commensurate stacking configurations.
Then, the real-space hopping terms were Fourier transformed, to calculate the effective interlayer coupling near the Dirac point, for every stacking configuration.
An additional Fourier transform in the reciprocal lattice vectors for the moire patterns led to the $W_m$ terms used in Eq.~(\ref{moireeq}) and defined in Eq.~(\ref{wexpression}).
The parameters used in our model Hamiltonian are:
\begin{equation}\label{eq:parameters}
\begin{array}{ll}
C_{0} = 0.01013~{\rm eV}, & \phi_{0} = 26.53^{\circ}~, \\
C_{z} = 0.00901~{\rm eV}, & \phi_{z} = -51.57^{\circ}~, \\
C_{AB} = 0.01134~{\rm eV}, & \phi_{xy} = 130.40^{\circ}~. \\
\end{array}
\end{equation}
The model also needs the following additional parameters:
\begin{eqnarray}
\begin{array}{ll}
\Delta_{0} = 0.010~{\rm eV}, & \epsilon = -0.017~, \\
v_{\rm F} = 1.1~{\rm nm}/{\rm fs}, &
\alpha_{\rm ee} = 1.0~,
\end{array}
\end{eqnarray}
where $\Delta_{0}$ is the magnitude of the average gap that can be introduced by moir\'e strains,~\cite{jung2015origin} $v_{\rm F}$ is the Fermi velocity in G/hBN, and $\alpha_{\rm ee} = e^{2} / (\bar{\epsilon} \hbar v_{\rm F})$ is a dimensionless coupling constant measuring the strength of electron-electron interactions, with $\bar{\epsilon}$ the average dielectric constant.

We point out that a different set of Hamiltonian parameters $u_{i}$, $\tilde{u}_{i}$, based on inversion symmetry considerations and a choice of origin, was also analyzed in the literature.~\cite{wallbank2013generic,wallbank2015moire}
The relation between the sets $u_{i}$, $\tilde{u}_{i}$ and $C_{\mu}$, $\phi_{\mu}$ is discussed in detail in Ref.~\onlinecite{jung_prb_2017}.

\subsection{Theory of the dielectric and loss functions for moir\'e superlattices}

The dielectric function $\epsilon_{{\bm G}, {\bm G}'}({\bm q}, \omega)$ relates the external potential $V_{\rm ext}({\bm G} + {\bm q}, \omega)$ applied to the electron system and the screened potential $V_{\rm sc}({\bm G} + {\bm q}, \omega)$, which results from the displacement of the carrier's charges, according to the relation
\begin{equation}
\sum_{{\bm G}'}\epsilon_{{\bm G}, {\bm G}'}({\bm q}, \omega) V_{\rm sc}({\bm G}' + {\bm q}, \omega) = V_{\rm ext}({\bm G} + {\bm q}, \omega)~.
\end{equation}
Here, ${\bm G}$ are reciprocal lattice vectors of the moir\'e pattern superlattice and ${\bm q}$ is a wave vector in the moir\'e Brillouin zone (mBZ).
Differently from a homogeneous system, the dielectric function is a matrix in the reciprocal lattice space,~\cite{Giuliani_and_Vignale} because the wave vector of the external potential is conserved only up to a reciprocal lattice vector of the mBZ.

Within the RPA, the dielectric function is related to the non-interacting density-density polarization function (i.e.~the Lindhard function) $\chi_{{\bm G},{\bm G}'}^{(0)}({\bm q},\omega)$ by~\cite{Giuliani_and_Vignale, neglect_hartree}
\begin{equation}\label{eq:RPA}
\epsilon_{{\bm G},{\bm G}'}({\bm q},\omega) = \delta_{{\bm G},{\bm G}'} - v_{{\bm G}}({\bm q}) \chi_{{\bm G},{\bm G}'}^{(0)}({\bm q},\omega)~,
\end{equation}
where $v_{{\bm G}}({\bm q}) = v({\bm q} + {\bm G})$ with $v(q) = 2\pi e^2/(\bar{\epsilon} q)$ is the 2D Fourier transform of the Coulomb potential.

The explicit expression for the Lindhard function is
\begin{equation} \label{eq:moireLindhard}
\begin{split}
&\chi^{(0)}_{{\bm G},{\bm G}'}({\bm q},\omega) =  \frac{2}{L^{2}}\sum_{{\bm k},n; {\bm k}', n'; \nu} \frac{n_{\rm F}(\varepsilon_{{\bm k},n,\nu}) - n_{\rm F}(\varepsilon_{{\bm k}',n',\nu})}{\hbar \omega + \varepsilon_{{\bm k},n,\nu} - \varepsilon_{{\bm k}',n',\nu} + i \eta}  \\
&\times  {\cal M}_{{\bm k}, n, \nu; {\bm k}', n', \nu}({\bm q} + {\bm G}) {\cal M}^\dagger_{{\bm k}, n, \nu; {\bm k}', n', \nu}({\bm q} + {\bm G}')~,
\end{split}
\end{equation}
where $L^2$ is the 2D electron system area, $n_{\rm F}(x)  = \{\exp[(x- \mu)/k_{\rm B} T] +1\}^{-1}$ is the Fermi-Dirac occupation factor at temperature $T$ and chemical potential $\mu$, and, finally,
\begin{equation}
{\cal M}_{{\bm k}, n, \nu; {\bm k}', n', \nu}({\bm q} + {\bm G}) \equiv \langle {\bm k},n,\nu | e^{-i({\bm q} + {\bm G}) \cdot {\bm r}} | {\bm k}',n',\nu\rangle~,
\end{equation}
$| {\bm k}, n, \nu \rangle$ being the eigenstate of the non-interacting Hamiltonian for wave vector ${\bm k}$ in the mBZ, band $n$, and principal valley $\nu$, and $\varepsilon_{{\bm k}, n, \nu}$ the corresponding eigenvalue.

The plasmon spectrum can be found, in principle, by solving for the roots $\omega = \omega_{\rm pl}({\bm q})$ of the ${\bm G} = 0$, ${\bm G}' = 0$ entry of the dielectric function matrix, in the wave vector ${\bm q}$ and \emph{complex} angular frequency $\omega$ space.
Assuming that the plasmon modes are long-lived, one can also solve for the roots with real angular frequency only, and then estimate the imaginary part.~\cite{fetter_walecka}
It is numerically more convenient, however, to calculate the so-called loss function
\begin{equation}
L({\bm q}, \omega) = - \Im m\{[\epsilon^{-1}]_{{\bm 0}, {\bm 0}}({\bm q}~,\omega)\}~,
\end{equation}
which is proportional to the probability of exciting the 2D electron system by applying a perturbation with wave vector ${\bm q}$ and angular frequency $\omega$, and is directly measured e.g.~via electron-energy-loss spectroscopy.~\cite{egerton_rpp_2009}
More details on the calculation of the loss function and the eigenvectors of the moir\'e Hamiltonian can be found in Ref.~\onlinecite{tomadin_prb_2014}.

\section{Results}
\label{sec:results}

\begin{figure}
\begin{overpic}[width=1.0\linewidth]{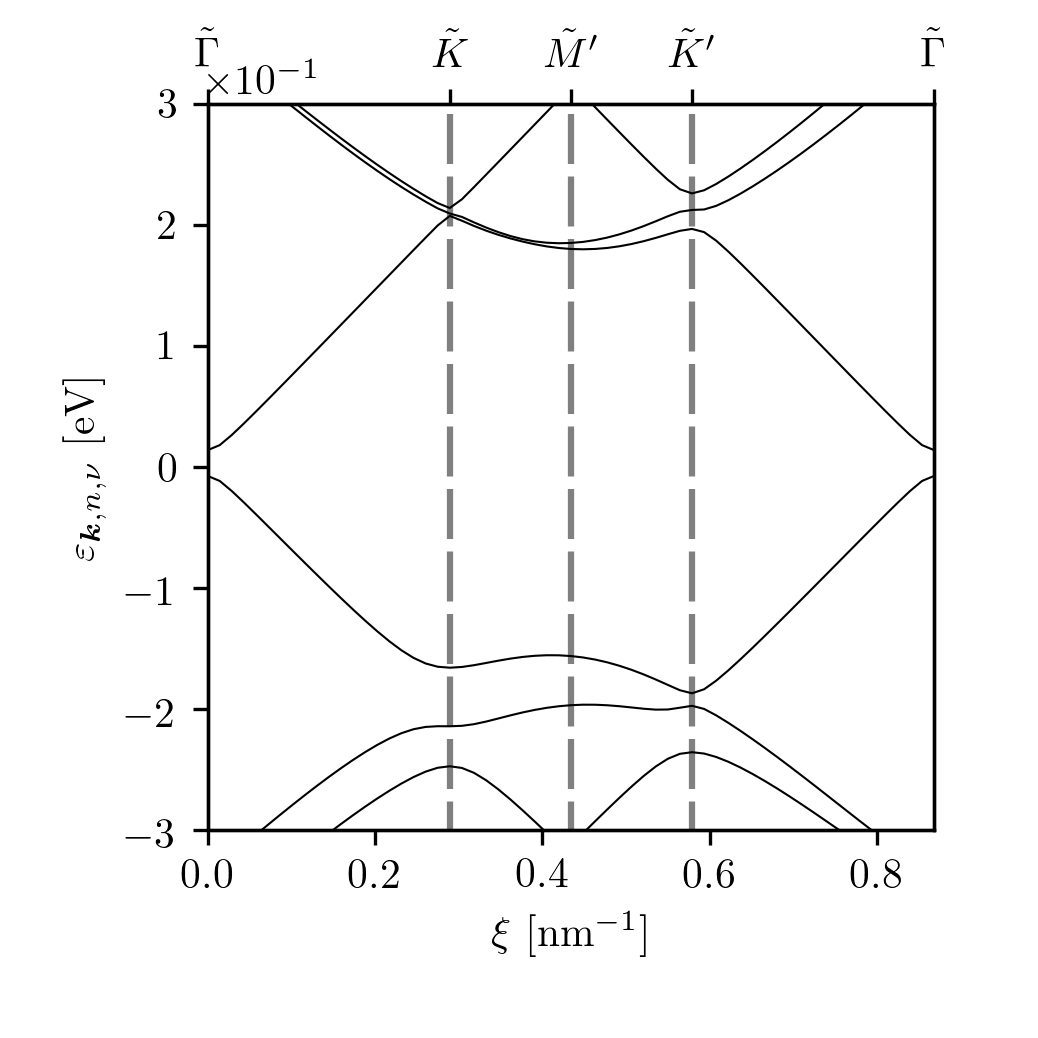}\put(2,55){}\end{overpic}
\caption{\label{fig:bands}
Graphene-hBN superlattice minibands along the $\tilde{\Gamma}$-$\tilde{K}$-$\tilde{M}'$-$\tilde{K}'$-$\tilde{\Gamma}$ direction in the mBZ,
for the set of parameters given in Eq.~\ref{eq:parameters}, corresponding to vanishing twist angle $\theta = 0$.
On the horizontal axis, the quantity $\xi$ indicates the total length along the path in the reciprocal space.}
\end{figure}

\begin{figure}
\begin{overpic}[width=1.0\linewidth]{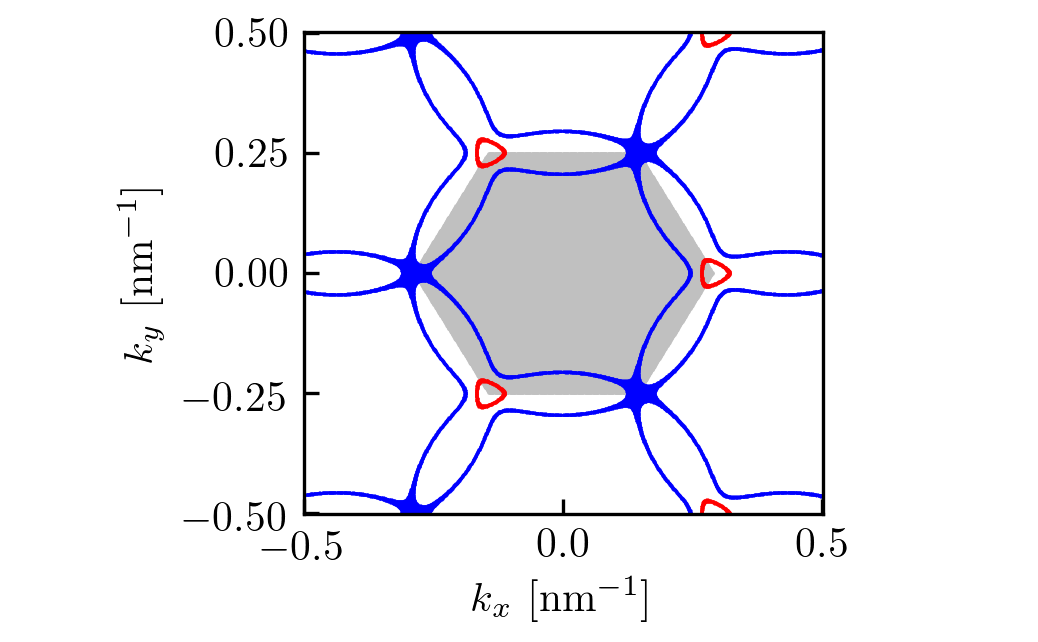}\put(2,55){}\end{overpic}
\caption{\label{fig:fermi_surfaces}
(Color online) Fermi surfaces of the minibands in Fig.~\ref{fig:bands} for two Fermi energies $\varepsilon_{\rm F} = -180~{\rm meV}$ (red) and $-215~{\rm meV}$ (blue).
The gray shaded area corresponds to the moir\'e superlattice Brillouin zone.
The Fermi surfaces are periodically repeated in the reciprocal space for clarity. }
\end{figure}

In this Section, we present our numerical results for the loss function and associated plasmon spectrum.

The dispersion of the electronic energies in the mBZ, dubbed ``minibands,'' is shown in Fig.~\ref{fig:bands} along a path in the mBZ around the $K$ valley of the original graphene Brillouin zone.
Two gaps are clearly visible at the energy of the Dirac point $\varepsilon = 0$ around the $\tilde{\Gamma}$ point of the mBZ and at $\varepsilon \simeq -200~{\rm meV}$ around the $\tilde{K}'$ point.
(Notice that, in the $K'$ valley, the points $\tilde{K}$ and $\tilde{K}'$ are exchanged.)
The flatness of the second band below the Dirac point is noteworthy and clearly visible along the path between the $\tilde{K}$ and $\tilde{K}'$ points.
To better appreciate this band's flatness, in Fig.~\ref{fig:fermi_surfaces} we plot the Fermi surfaces for two different Fermi energies.
The bottom of the first band below the Dirac point consists of one sharp minimum around the $\tilde{K}'$ point, parabolic in shape but strongly anisotropic.
The maximum of the second band, instead, is located at the $\tilde{M}$ point but, in a small energy range $\lesssim 5~{\rm meV}$, expands around the $\tilde{K}'$ point, roughly in the shape of a three-blade propeller.
Eventually, the tips of the blade-like shapes touch at the $\tilde{K}$ points and merge, yielding a familiar-looking but distorted hexagonal Fermi surface.

\begin{figure}
\begin{overpic}[width=1.0\linewidth]{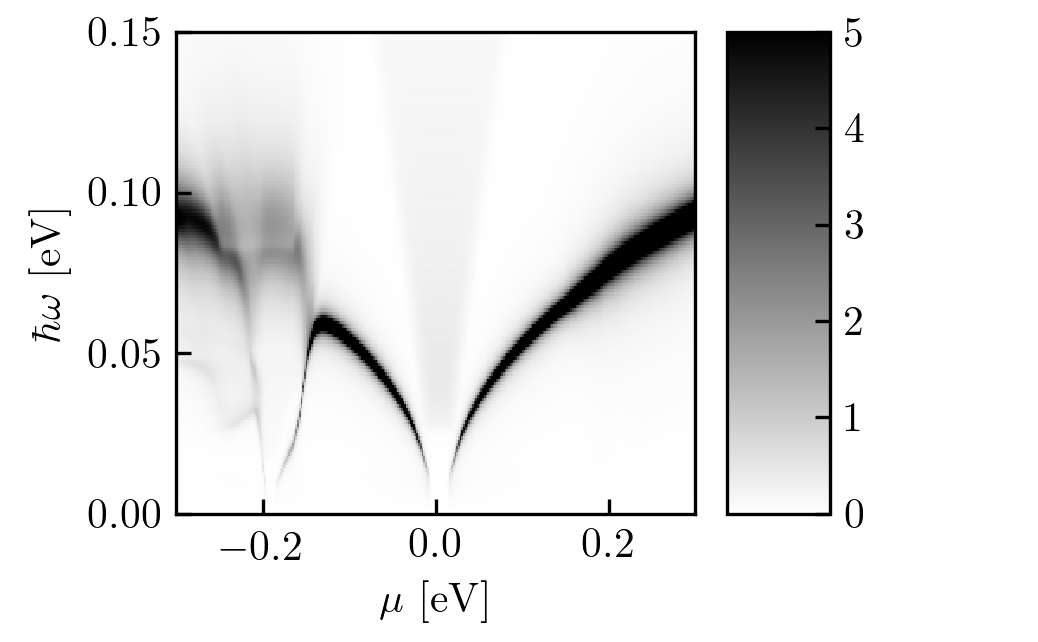}\put(2,55){(a)}\end{overpic}
\begin{overpic}[width=1.0\linewidth]{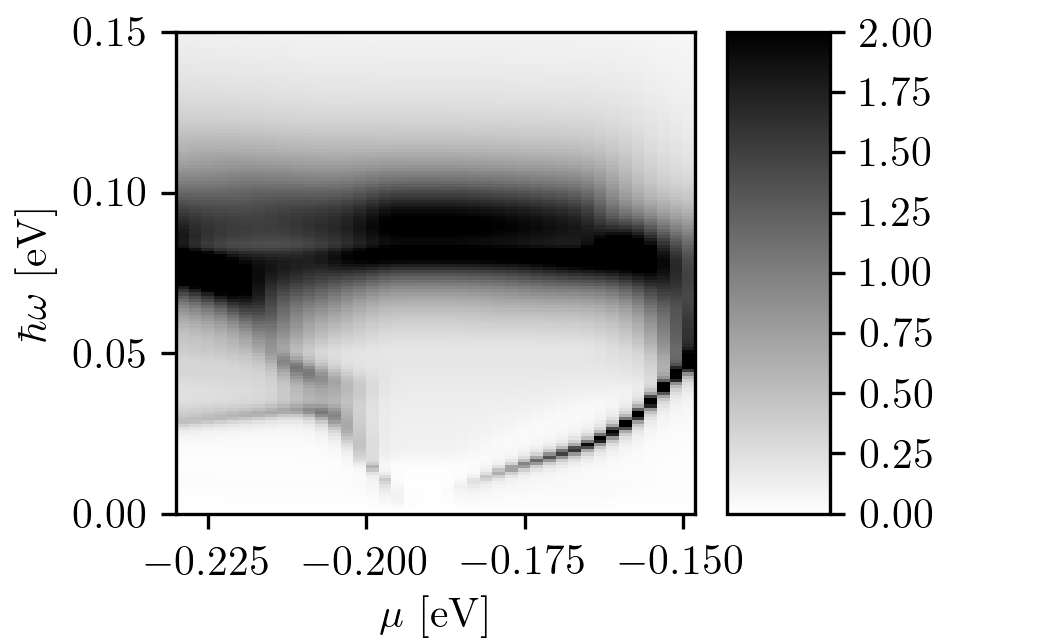}\put(2,55){(b)}\end{overpic}
\caption{\label{fig:loss_vs_mu}
A 2D density plot of the RPA loss function $L({\bm q}, \omega)$, as a function of the chemical potential $\mu$ and excitation energy $\hbar \omega$, for the parameters given in Eq.~\ref{eq:parameters}.
The calculation is performed for a fixed wave vector of length $q = 0.02~{\rm nm}^{-1}$ along the $\tilde{M}$-$\tilde{\Gamma}$ direction.
Panel (b) shows a magnification of (a) around the gap at $\mu \simeq -200~{\rm meV}$ (cfr.~the minibands in Fig.~\ref{fig:bands}).
In both panels, and in the following figures as well, the range of the monochrome shades has been truncated to improve the visibility of the less intense features.
For comparison, the graph of the loss function at fixed chemical potential is shown in Fig.~\ref{fig:dielectric_function}(a).}
\end{figure}

Fig.~\ref{fig:loss_vs_mu} contains the main results of this Section.
It shows the loss function, at fixed wave vector, in a large chemical potential range.
Above the Dirac point, we identify a single plasmon branch, almost unperturbed by the moir\'e potential.
We emphasize that the spectral broadening of the plasmon branch, i.e.~the width of the peak as a function of $\omega$, cannot be readily estimated from this density plot, because the extent of the monochrome shades has been truncated to improve the visibility of the less intense features.
Around the Dirac point, the existence of the gap manifests as a forbidden band for the plasmon propagation, i.e.~a region where plasmon branches are not supported.
Moreover, inter-band transitions across the gap contribute a continuum of excitations which has the shape of an inverted, truncated cone.

The profile of the loss function below the Dirac point is dramatically different.
As the chemical potential becomes more negative, the graphene's plasmon branch first grows in energy and then bends abruptly to reach zero energy at the gap located around $\mu \simeq - 200~{\rm meV}$.
Below the gap, a plasmon branch rises again.
This is an instance of the \emph{plasmon morphing} phenomenon that was introduced and discussed in Ref.~\onlinecite{tomadin_prb_2014}.
Across the gap, inter-band transitions contribute a thick continuum around $\hbar \omega \lesssim 10~{\rm meV}$.
The location and extent of this continuum can be understood by looking at the Fermi surfaces just below and above the gap, shown in Fig.~\ref{fig:fermi_surfaces}, which support a large number of electronic transitions with almost arbitrary wave vector in a restricted energy range.
Most interestingly, below the gap and as the chemical potential changes, more than one plasmon branch and an apparent ``avoided crossing'' appear, suggesting that these branches correspond to coupled modes.

\begin{figure}
\begin{overpic}[width=1.0\linewidth]{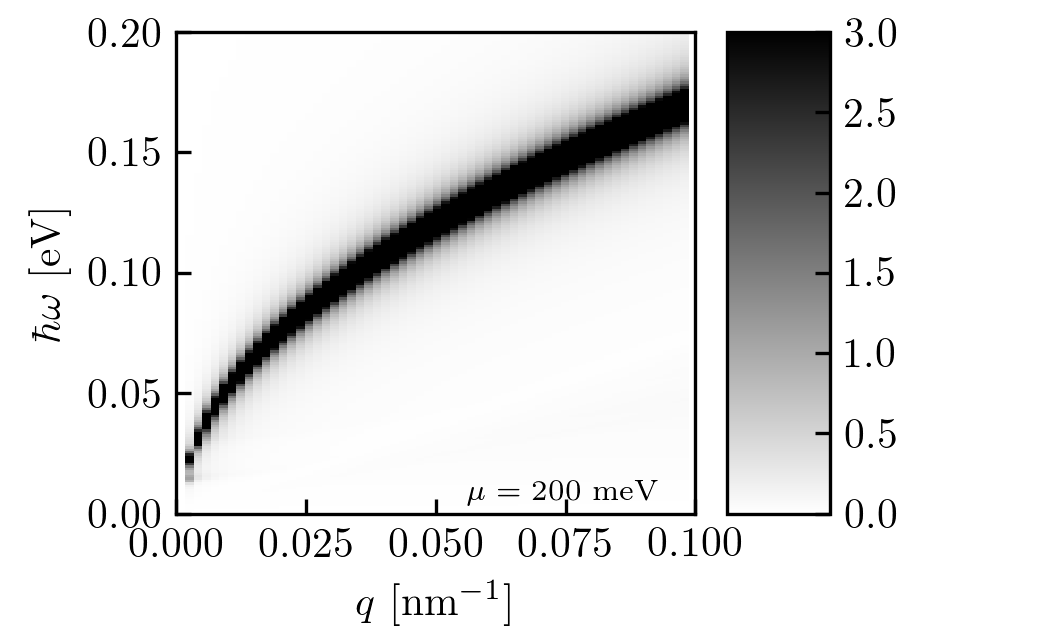}\put(2,55){(a)}\end{overpic}
\begin{overpic}[width=1.0\linewidth]{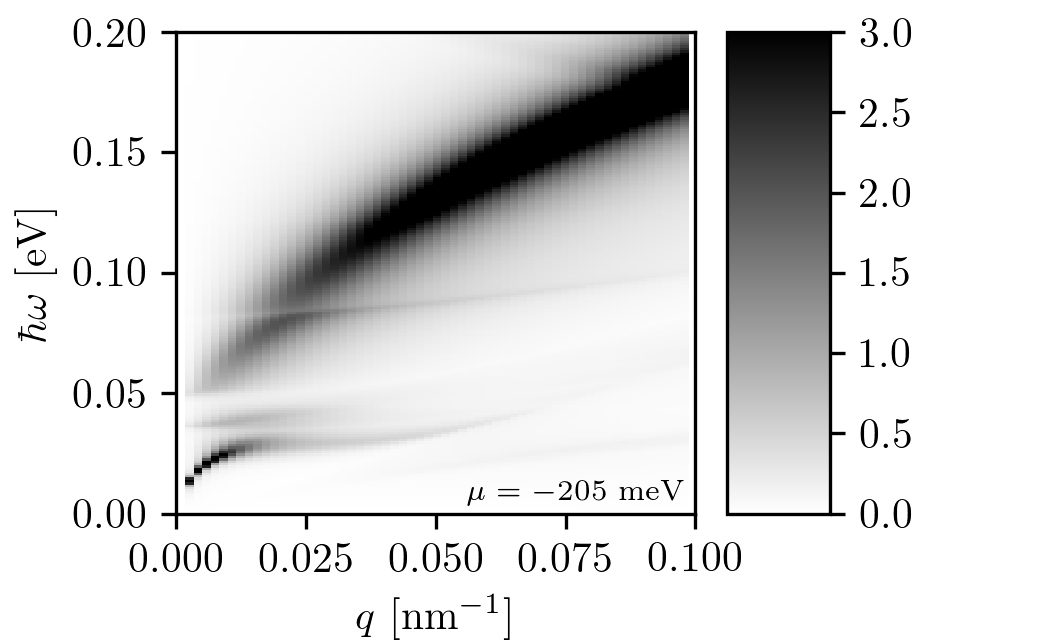}\put(2,55){(b)}\end{overpic}
\caption{\label{fig:loss_vs_q}
A 2D density plot of the RPA loss function $L({\bm q}, \omega)$, as a function of the wave vector length $q$ and the excitation energy $\hbar \omega$.
The wave vector is taken along the $\tilde{M}$-$\tilde{\Gamma}$ direction.
Results for positive $\mu = 200~{\rm meV}$ and negative $\mu = -205~{\rm meV}$ chemical potentials are shown in (a) and (b), respectively. }
\end{figure}

\begin{figure}
\begin{overpic}[width=1.0\linewidth]{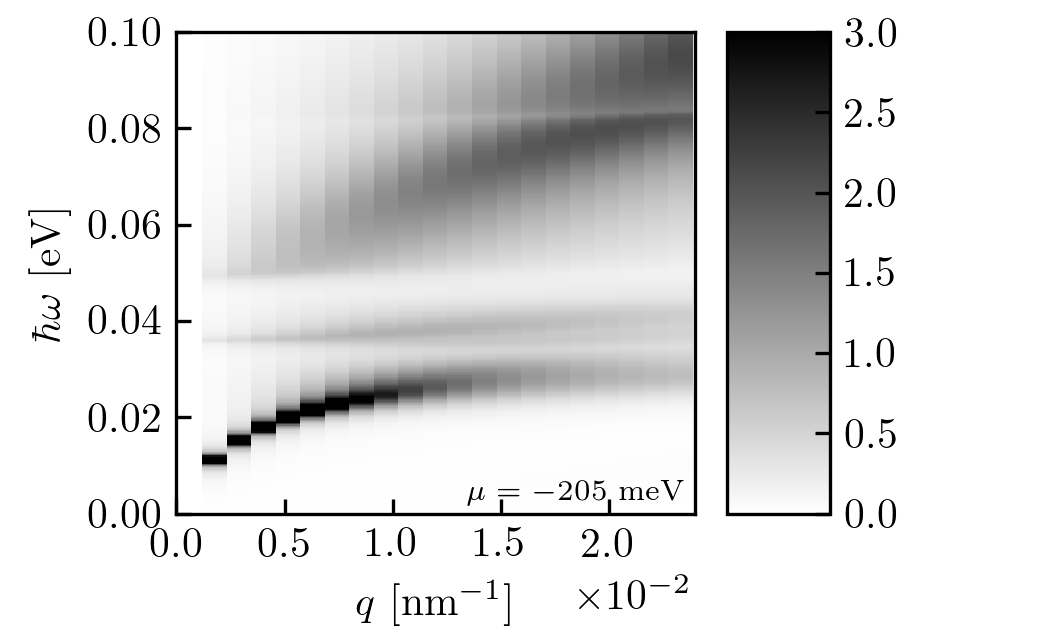}\put(2,55){(a)}\end{overpic}
\begin{overpic}[width=1.0\linewidth]{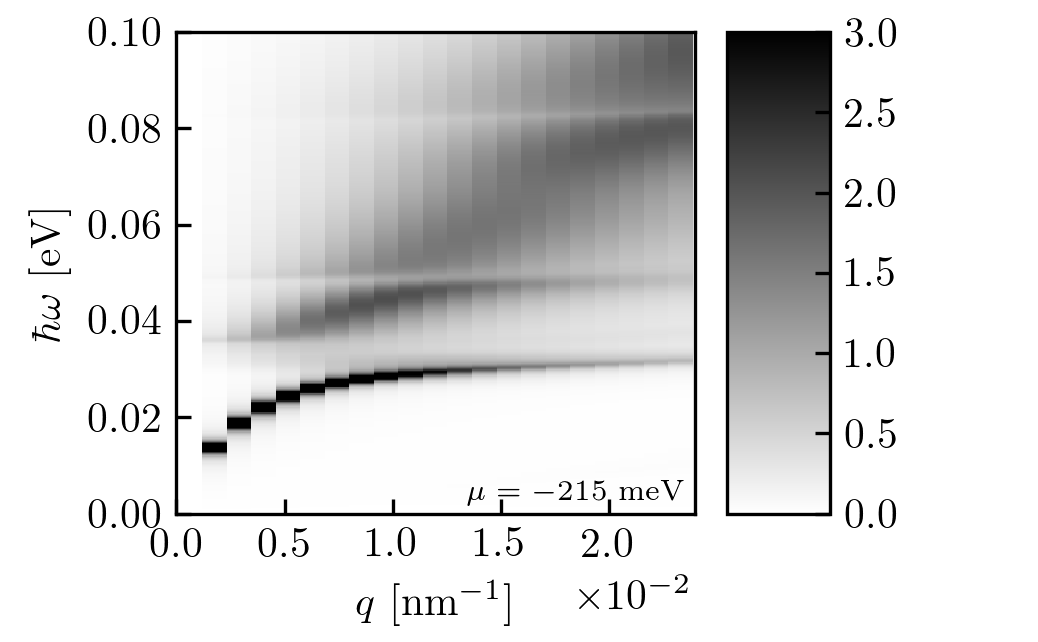}\put(2,55){(b)}\end{overpic}
\caption{\label{fig:loss_vs_q_zoom}
As in Fig.~\ref{fig:loss_vs_q}, but in a smaller wave vector range around the origin of the reciprocal space.
Results for two negative chemical potentials $\mu = -205~{\rm meV}$ and $-215~{\rm meV}$, slightly lower than the gap (cfr.~the minibands in Fig.~\ref{fig:bands}), are shown in (a) and (b), respectively. }
\end{figure}

To better appreciate the asymmetry of the plasmon spectrum above and below the Dirac point, in Fig.~\ref{fig:loss_vs_q} we show the loss function as a function of the wave vector, for two values of the chemical potential.
Above the Dirac point the plasmon's dispersion is almost unperturbed by the moir\'e potential.
On the contrary, close to the gap at $\mu \simeq -200~{\rm meV}$, the low-energy dispersion ($\hbar \omega \lesssim 100~{\rm meV}$) is fractured into several branches with variable intensity, and recovers its almost unperturbed profile only at larger energies.
Fig.~\ref{fig:loss_vs_q_zoom} focuses on the low-energy dispersion for two chemical potentials close to the gap.
The dispersion is very similar, thus showing that the features discussed here are robust and do not depend on the specific value of the chemical potential.
At very low energies ($\hbar \omega \lesssim 30~{\rm meV}$) a well-defined branch rises with $q$ and then flattens out, giving way to a continuum band of excitations, peaked around the unperturbed dispersion.
The continuum band features a thin abrupt fracture around $\hbar \omega \simeq 50~{\rm meV}$, above which another continuum band appears.

\begin{figure}
\begin{overpic}[width=1.0\linewidth]{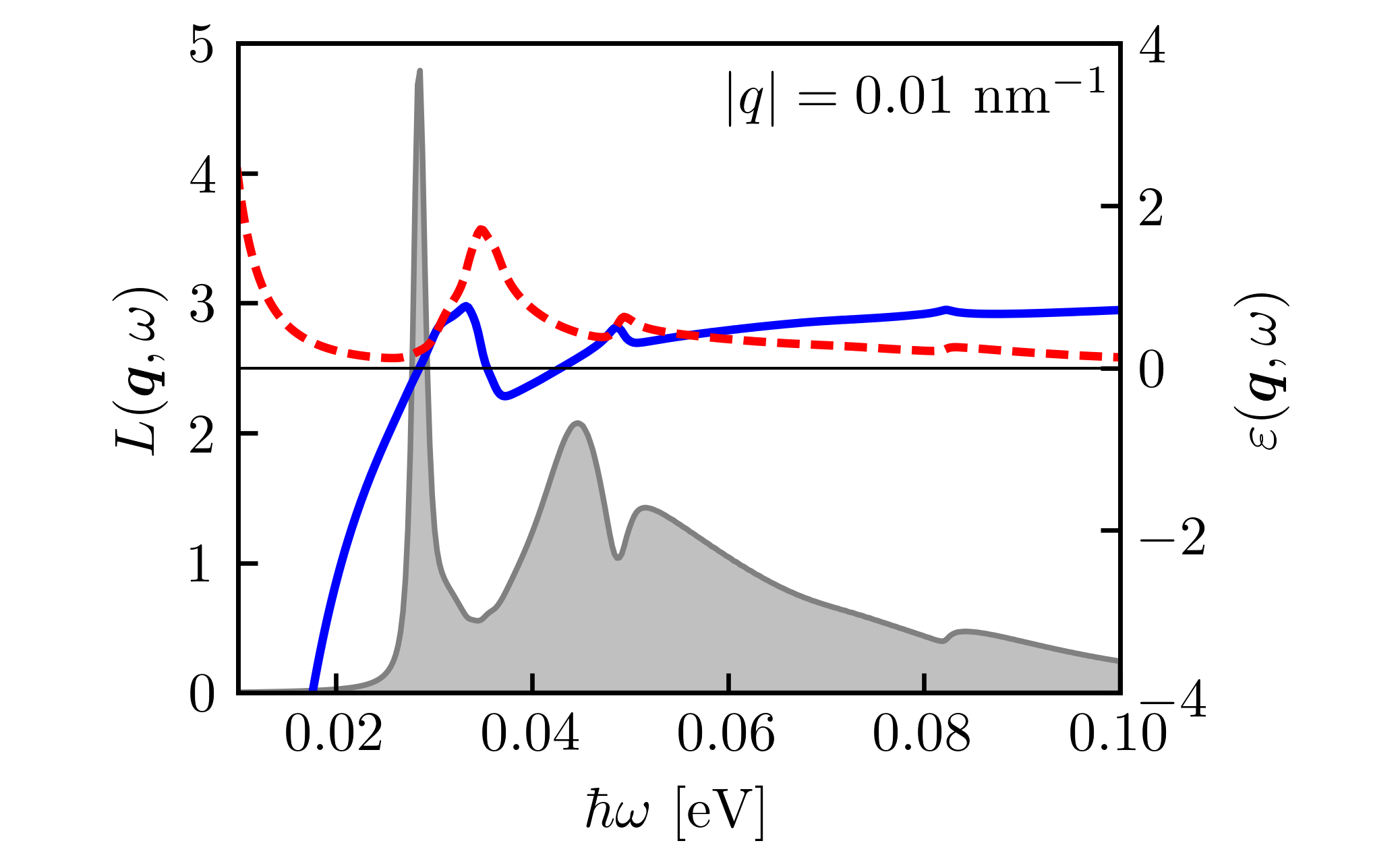}\put(2,55){(a)}\end{overpic}
\begin{overpic}[width=1.0\linewidth]{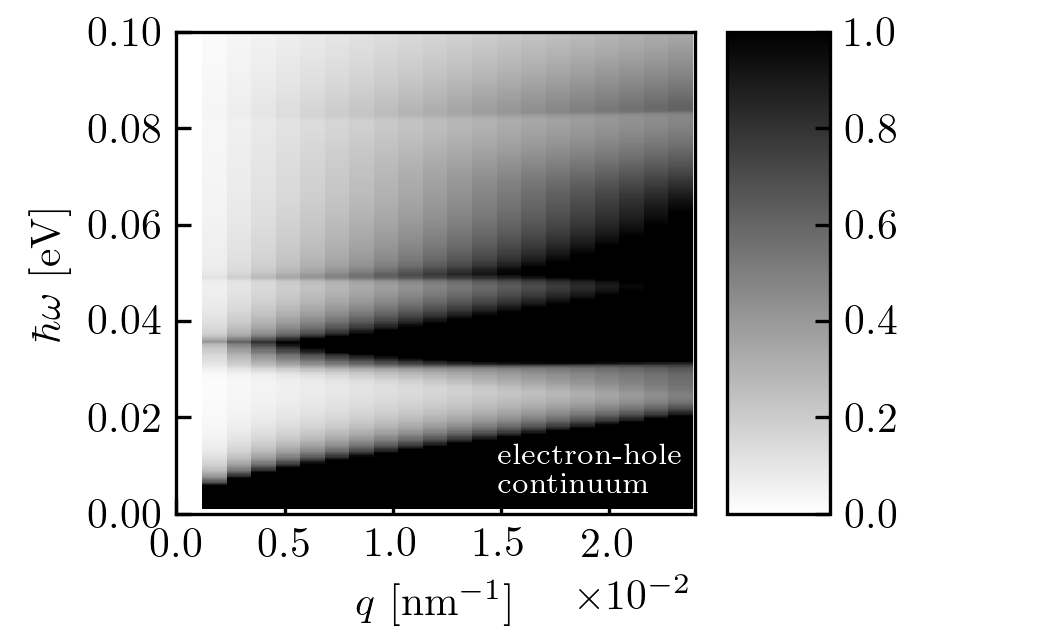}\put(2,55){(b)}\end{overpic}
\caption{\label{fig:dielectric_function}
(Color online) (a) The RPA loss function $L({\bm q}, \omega)$ (shaded area), the real (solid line), and the imaginary (dashed line) part of the dielectric function $\epsilon({\bm q}, \omega)$, as functions of the transition energy $\hbar \omega$, at fixed wave vector length $q = 0.01~{\rm nm}^{-1}$.
The values of the loss (dielectric) function are reported on the left (right) vertical axis.
(b) A 2D density plot of the imaginary part of the dielectric function, as a function of the wave vector length $q$ and the excitation energy $\hbar \omega$.
In both panels, the wave vector is taken along the $\tilde{M}-\tilde{\Gamma}$ direction and the chemical potential is $\mu = -215~{\rm meV}$, as in Fig.~\ref{fig:loss_vs_q}(b). }
\end{figure}

To guide the interpretation of these spectral features, in Fig.~\ref{fig:dielectric_function}(a) we juxtapose the loss function and the real and imaginary parts of the dielectric function.
Two sharp zeros of the real part of the dielectric function are present where the loss function has its maxima.
This shows that the maxima are indeed \emph{collective modes}, and can be interpreted as branches of the plasmon dispersion.
Between the maxima, a peak of the imaginary part of the dielectric function signals the existence of a continuum of electronic excitations, which separates the two branches.
A more complete picture is obtained by looking at Fig.~\ref{fig:dielectric_function}(b), which shows the imaginary part of the dielectric function, i.e.~the electron-hole (e-h) continuum.
In unperturbed graphene, at excitation energies which are small compared to the chemical potential, the e-h continuum consists of intra-band excitations below the ``light-cone'' $\omega = v_{\rm F} q$.
Here, instead, it consists of a fractured domain, which includes bands extending horizontally with sharp bottom edges.
Again, the origin of these bands can be qualitatively understood by referring to the large Fermi surfaces shown in Fig.~\ref{fig:fermi_surfaces}.
More precisely, elongated structures in the Fermi surface grant a support for e-h transitions with variable wave vector and constant energy, which coalesce into the horizontal bands of Fig.~\ref{fig:dielectric_function}(b).
The dispersion of the plasmon modes cannot penetrate the intra-band e-h continuum, as is the case in unperturbed graphene, where the plasmon dispersion is tangent to the light-cone.
Thus, the dispersion bends and follows the bottom edge of the e-h bands.
Different branches effectively avoid each other, as if they were coupled modes, because they stem from the same set of electronic excitations.

The results that we have presented above hold for nearly-aligned graphene and hBN layers, where $\theta \simeq 0$, because they crucially depend on the existence of the gap in the electronic dispersion at the $\tilde{K}'$ point of the Brillouin zone around $\mu \simeq -200~{\rm meV}$.
However, we have verified that the plasmon spectrum is asymmetric for a larger angle $\theta = 2^{\circ}$ as well, although in a less dramatic fashion than displayed in Fig.~\ref{fig:loss_vs_mu}.
Since the exact angular dependence of the parameter $\Delta_{0}$ is not known analytically, and experimental reports of the gap magnitude are not in agreement~\cite{hunt2013massive,florida_2014,wang2016direct,woods2014commensurate,suyongjung2018}, we have used both $\Delta_{0} = 0$ and $\Delta_{0} = 10~{\rm meV}$ in the calculations.

\section{Summary and conclusions}
\label{sec:summary}

In this work we have analyzed the plasmon spectrum of a heterostructure composed of two nearly-aligned layers of graphene and hexagonal boron nitride (hBN).
We have used a continuum-model effective Hamiltonian to obtain the dispersion relation of graphene's carriers in the heterostructure, which is different from that of isolated graphene because the hBN layer generates a periodic moir\'e potential for the carriers.
We have discussed in detail the symmetry of the moir\'e potential and the relation between different formal representations of its functional form in real and sublattice space.
The parameters of the moir\'e potential have been derived using a framework~\cite{jung_prb_2017} which combines symmetry considerations with input from {\it ab initio} calculations.
The electronic dispersion obtained with the continuum model consists of several minibands in the moir\'e superlattice Brillouin zones, centered at the $K$ and $K'$ points of the Brillouin zone of pristine graphene, which shift in energy as the twist angle between the layers is varied.
At vanishing twist angle between the layers, a gap is present about $200~{\rm meV}$ below the Dirac point.
We have numerically calculated the dielectric function and the loss function taking into account electronic transitions between minibands in the
moir\'e Brillouin zones and electron-electron interactions at the level of the random phase approximation (RPA).

In conclusion, our calculations demonstrate a dramatic asymmetry of the plasmon dispersion at positive and negative chemical potential.
This observation is potentially very relevant to establish the ideal working point of a graphene/hBN heterostructure as a two-dimensional platform for tunable, low-loss plasmonics.
Moreover, around the gap below the Dirac point, the plasmon spectrum features several branches which appear as a result of a fractured electron-hole continuum due to the inter-band transitions between closely-spaced minibands with almost flat dispersion.
Given the richness of the available band dispersion in graphene-based and, in general, in van der Waals heterostructures, our findings could be useful to guide further exploration of the non-trivial connection between the electronic and plasmonic dispersion in these systems.

\acknowledgments

This work was supported by the Korean NRF through the grant number 2016R1A2B4010105 (J.J.) and the European Union's Horizon 2020 research and innovation programme under grant agreement No. 785219 - ``GrapheneCore2'' (A.T. and M.P.).
Financial support from Consiglio Nazionale delle Ricerche (CNR) in the framework of the agreements on scientific collaborations between CNR and NRF (Korea) is also acknowledged in the initial stages of this work.


\begin{thebibliography}{88}

\bibitem{novoselov2004electric}
K.S. Novoselov, A.K. Geim, S.V. Morozov, D. Jiang, Y. Zhang, S.V. Dubonos, I.V. Grigorieva, and A.A. Firsov, \href{https://doi.org/10.1126/science.1102896}{Science {\bf 306}, 666 (2004)}.

\bibitem{novoselov2005two}
K.S. Novoselov, D. Jiang, F. Schedin, T.J. Booth, V.V. Khotkevich, S.V. Morozov, and A.K. Geim, \href{https://doi.org/10.1073/pnas.0502848102}{Proc. Natl. Acad. Sci. U.S.A. {\bf 102}, 10451 (2005)}.

\bibitem{geim2007rise}
A.K. Geim and K.S. Novoselov, \href{https://doi.org/10.1038/nmat1849}{Nature Mater. {\bf 6}, 183 (2007)}.

\bibitem{neto2009electronic}
A.H. Castro Neto, F. Guinea, N.M.R. Peres, K.S. Novoselov, and A.K. Geim, \href{https://doi.org/10.1103/RevModPhys.81.109}{Rev. Mod. Phys. {\bf 81}, 109 (2009)}.

\bibitem{sarma2011electronic}
S. Das Sarma, S. Adam, E.H. Hwang, and E. Rossi, \href{https://doi.org/10.1103/RevModPhys.83.407}{Rev. Mod. Phys. {\bf 83}, 407 (2011)}.

\bibitem{dean2010boron}
C.R. Dean, A.F. Young, I. Meric, C. Lee, L. Wang, S. Sorgenfrei, K. Watanabe, T. Taniguchi, P. Kim, K.L. Shepard, and J. Hone, \href{https://doi.org/10.1038/nnano.2010.172}{Nature Nanotechnol. {\bf 5}, 722 (2010)}.

\bibitem{geim2013van}
A.K. Geim and I.V. Grigorieva, \href{https://doi.org/10.1038/nature12385}{Nature {\bf 499}, 419 (2013)}.

\bibitem{xue2011scanning}
J. Xue, J. Sanchez-Yamagishi, D. Bulmash, P. Jacquod, A. Deshpande, K. Watanabe, T. Taniguchi, P. Jarillo-Herrero, and B.J. LeRoy, \href{https://doi.org/10.1038/nmat2968} {Nature Mater. {\bf 10}, 282 (2011)}.

\bibitem{du2009fractional}
X. Du, I. Skachko, F. Duerr, A. Luican, and E.Y. Andrei, \href{https://doi.org/10.1038/nature08522}{Nature {\bf 462}, 192 (2009)}.

\bibitem{bolotin2009observation}
K.I. Bolotin, F. Ghahari, M.D. Shulman, H.L. Stormer, and P. Kim, \href{https://doi.org/10.1038/nature08582}{Nature {\bf 462}, 196 (2009)}.

\bibitem{elias2011dirac}
D.C. Elias, R.V. Gorbachev, A.S. Mayorov, S.V. Morozov, A.A. Zhukov, P. Blake, L.A. Ponomarenko, I.V. Grigorieva, K.S. Novoselov, F. Guinea, and A.K. Geim, \href{https://doi.org/10.1038/nphys2049}{Nature Phys. {\bf 7}, 701 (2011)}.

\bibitem{gorbachev2012strong}
R.V. Gorbachev, A.K. Geim, M.I. Katsnelson, K.S. Novoselov, T. Tudorovskiy, I.V. Grigorieva, A.H. MacDonald, S.V. Morozov, K. Watanabe, T. Taniguchi, and L.A. Ponomarenko, \href{https://doi.org/10.1038/nphys2441}{Nature Phys. {\bf 8}, 896 (2012)}.

\bibitem{dean2013hofstadter}
C.R. Dean, L. Wang, P. Maher, C. Forsythe, F. Ghahari, Y. Gao, J. Katoch, M. Ishigami, P. Moon, M. Koshino, T. Taniguchi, K. Watanabe, K.L. Shepard, J. Hone, and P. Kim, \href{https://doi.org/10.1038/nature12186}{Nature {\bf 497}, 598 (2013)}.

\bibitem{ponomarenko2013cloning}
L.A. Ponomarenko, R.V. Gorbachev, G.L. Yu, D.C. Elias, R. Jalil, A.A. Patel, A. Mishchenko, A.S. Mayorov, C.R. Woods, J.R. Wallbank, M. Mucha-Kruczynski, B.A. Piot, M. Potemski, I.V. Grigorieva, K.S. Novoselov, F. Guinea, V.I. Fal'ko, and A.K. Geim, \href{https://doi.org/10.1038/nature12187}{Nature {\bf 497}, 594 (2013)}.

\bibitem{hunt2013massive}
B. Hunt, J.D. Sanchez-Yamagishi, A.F. Young, M. Yankowitz, B.J. LeRoy, K. Watanabe, T. Taniguchi, P. Moon, M. Koshino, P. Jarillo-Herrero, and R.C. Ashoori, \href{https://doi.org/10.1126/science.1237240}{Science {\bf 340}, 1427 (2013)}.

\bibitem{amet2013insulating}
F. Amet, J.R. Williams, K. Watanabe, T. Taniguchi, and D. Goldhaber-Gordon, \href{https://doi.org/10.1103/PhysRevLett.110.216601}{Phys. Rev. Lett. {\bf 110}, 216601 (2013)}.

\bibitem{florida_2014}
Z.-G. Chen, Z. Shi, W. Yang, X. Lu, Y. Lai, H. Yan, F. Wang, G. Zhang and Z. Li, \href{https://doi.org/10.1038/ncomms5461}{Nature Commun. {\bf 5}, 4461 (2014)}.

\bibitem{wang2016direct}
E. Wang, X. Lu, S. Ding, W. Yao, M. Yan, G. Wan, K. Deng, S. Wang, G. Chen, L. Ma, J. Jung, A.V. Fedorov, Y. Zhang, G. Zhang, and S. Zhou, \href{https://doi.org/10.1038/nphys3856}{Nature Phys. {\bf 12}, 1111 (2016)}.

\bibitem{woods2014commensurate}
C.R. Woods, L. Britnell, A. Eckmann, R.S. Ma, J.C. Lu, H.M. Guo, X. Lin, G.L. Yu, Y. Cao, R.V. Gorbachev, A.V. Kretinin, J. Park, L.A. Ponomarenko, M.I. Katsnelson, Y.N. Gornostyrev, K. Watanabe, T. Taniguchi, C. Casiraghi, H.-J. Gao, A.K. Geim, and K.S. Novoselov, \href{https://doi.org/10.1038/nphys2954}{Nature Phys. {\bf 10}, 451 (2014)}.

\bibitem{suyongjung2018}
H.S. Kim et al., to be submitted.

\bibitem{jung2015origin}
J. Jung, A.M. DaSilva, A.H. MacDonald, and S. Adam, \href{https://doi.org/10.1038/ncomms7308}{Nature Commun. {\bf 6}, 6308 (2015)}.

\bibitem{sanjose2014spontaneous}
P. San-Jose, A. Guti\'errez-Rubio, M. Sturla, and F. Guinea, \href{https://doi.org/10.1103/PhysRevB.90.075428}{Phys. Rev. B {\bf 90}, 75428 (2014)}.

\bibitem{dasilva2015transport}
A.M. DaSilva, J. Jung, S. Adam, and A.H. MacDonald, \href{https://doi.org/10.1103/PhysRevB.91.245422}{Phys. Rev. B {\bf 91}, 245422 (2015)}.

\bibitem{jung_prb_2017}
J. Jung, E. Laksono, A.M. DaSilva, A.H. MacDonald, M. Mucha-Kruczy\'nski, and S. Adam, \href{https://doi.org/10.1103/PhysRevB.96.085442}{Phys. Rev. B {\bf 96}, 085442 (2017)}.

\bibitem{bonaccorso_naturephoton_2010}
F. Bonaccorso, Z. Sun, T. Hasan, and A. Ferrari, \href{https://doi.org/10.1038/NPHOTON.2010.186}{Nature Photon. {\bf 4}, 611 (2010)}.

\bibitem{grigorenko_naturephoton_2012}
A. Grigorenko, M. Polini, and K. Novoselov, \href{https://doi.org/10.1038/nphoton.2012.262}{Nature Photon. {\bf 6}, 749 (2012)}.

\bibitem{woessner_naturemater_2015}
A. Woessner, M.B. Lundeberg, Y. Gao, A. Principi, P. Alonso-Gonz\'alez, M. Carrega, K. Watanabe, T. Taniguchi, G. Vignale, M. Polini, J. Hone, R. Hillenbrand, and F.H.L. Koppens, \href{https://doi.org/10.1038/NMAT4169}{Nature Mater. {\bf 14}, 421 (2015)}.

\bibitem{fei_nature_2012}
Z. Fei, A.S. Rodin, G.O. Andreev, W. Bao, A.S. McLeod, M. Wagner, L.M. Zhang, Z. Zhao, M. Thiemens, G. Dominguez, M.M. Fogler, A.H. Castro Neto, C.N. Lau, F. Keilmann, and D.N. Basov, \href{https://doi.org/10.1038/nature11253}{Nature {\bf 487}, 82 (2012)}.

\bibitem{chen_nature_2012}
J. Chen, M. Badioli, P. Alonso-Gonz\'alez, S. Thongrattanasiri, F. Huth, J. Osmond, M. Spasenovi\'c, A. Centeno, A. Pesquera, P. Godignon, A. Zurutuza Elorza, N. Camara, F.J. Garc\'ia de Abajo, R. Hillenbrand, and F.H.L. Koppens, \href{https://doi.org/10.1038/nature11254}{Nature {\bf 487}, 77 (2012)}.

\bibitem{ni_naturemater_2015}
G.X. Ni, H. Wang, J.S. Wu, Z. Fei, M.D. Goldflam, F. Keilmann, B. \"Ozyilmaz, A.H. Castro Neto, X.M. Xie, M.M. Fogler, and D.N. Basov, \href{https://doi.org/10.1038/nmat4425}{Nature Mater. {\bf 14}, 1217 (2015)}.

\bibitem{ni_nature_2018}
G.X. Ni, A.S. McLeod, Z. Sun, L. Wang, L. Xiong, K.W. Post, S.S. Sunku, B.-Y. Jiang, J. Hone, C.R. Dean, M.M. Fogler, and D.N. Basov, \href{https://doi.org/10.1038/s41586-018-0136-9}{Nature {\bf 557}, 530 (2018)}.

\bibitem{basov_science_2016}
D.N. Basov, M.M. Fogler, and F.J. Garc\'ia de Abajo, \href{https://doi.org/10.1126/science.aag1992}{Science {\bf 354}, aag1992 (2016)}.

\bibitem{low_naturemater_2017}
T. Low, A. Chaves, J. Caldwell, A. Kumar, N. Fang, P. Avouris, T. Heinz, F. Guinea, L. Martin-Moreno, and F. Koppens, \href{https://doi.org/10.1038/NMAT4792}{Nature Mater. {\bf 16}, 182 (2017)}.

\bibitem{tomadin_prb_2014}
A. Tomadin, F. Guinea, and M. Polini, \href{https://doi.org/10.1103/PhysRevB.90.161406}{Phys. Rev. B {\bf 90}, 161406(R) (2014)}.

\bibitem{jung2014ab}
J. Jung, A. Raoux, Z. Qiao, and A.H. MacDonald, \href{https://doi.org/10.1103/PhysRevB.89.205414}{Phys. Rev. B {\bf 89}, 205414 (2014)}.

\bibitem{Giuliani_and_Vignale}
G.F. Giuliani and G. Vignale, \href{https://doi.org/10.1080/00107510903194710} {\it Quantum Theory of the Electron Liquid} (Cambridge University Press, Cambridge, 2005).

\bibitem{footnote_phi}
In Ref.~\onlinecite{jung_prb_2017}, the notation $\varphi_{AB} = \pi / 6 - \phi_{xy}$ was used.

\bibitem{wallbank2013generic}
J.R. Wallbank, A.A. Patel, M. Mucha-Kruczy\'nski, A.K. Geim, and V.I. Fal'ko, \href{https://doi.org/10.1103/PhysRevB.87.245408}{Phys. Rev. B {\bf 87}, 245408 (2013)}.

\bibitem{wallbank2015moire}
J.R. Wallbank, M. Mucha-Kruczy\'nski, X. Chen, and V.I. Fal'ko, \href{https://doi.org/10.1002/andp.201400204}{Ann. Phys. (Berlin) {\bf 527}, 359 (2015)}.

\bibitem{neglect_hartree}
In this work, we neglect Hartree corrections to the Lindhard function, which in general are necessary to formulate the RPA of inhomogeneous systems.

\bibitem{fetter_walecka}
A.L. Fetter and J.D. Walecka, \href{https://isbnsearch.org/isbn/0486428273}{\it Quantum Theory of Many-Particle Systems} (Dover Publications, 2003).

\bibitem{egerton_rpp_2009}
R.F. Egerton, \href{https://doi.org/10.1088/0034-4885/72/1/016502}{Rep. Prog. Phys. {\bf 72}, 016502 (2009)}.

\end{thebibliography}
\end{document}